\documentclass[apjl]{emulateapj}
\usepackage{mathptmx}

\newcommand{\ltsim}{\raisebox{-.5ex}{$\;\stackrel{<}{\sim}\;$}}
\newcommand{\gtsim}{\raisebox{-.5ex}{$\;\stackrel{>}{\sim}\;$}}
\newcommand{\kms}{\ifmmode {\rm km\ s}^{-1} \else km s$^{-1}$\fi}
\newcommand{\lledd}{$L/L_{\rm Edd}$}

\newcommand{\mbh}{$M_{\rm BH}$}
\newcommand{\et}{et al.\ }
\newcommand{\xray}{\hbox{X-ray}}

\newcommand{\nh}{$N_{\rm H}$}

\newcommand{\hb}{H$\beta$}
\newcommand{\xmm}{{\sl XMM-Newton}}
\newcommand{\chandra}{{\sl Chandra}}
\newcommand{\geminiN}{{\sl Gemini-North}}

\usepackage{epstopdf}

\journalinfo{The Astrophysical Journal, ???:L??--L??, 2010 ??? ??}
\slugcomment{Received 2010 August 13; accepted 2010 September 8}

\shortauthors{SHEMMER ET AL.}
\shorttitle{WLQS: HIGH ACCRETION RATES OR ANEMIC BLRS?}

\begin{document}
\title{Weak-Line Quasars at High Redshift: Extremely High Accretion Rates or \\ Anemic Broad-Line Regions?}

\author{
Ohad~Shemmer,\altaffilmark{1}
Benny~Trakhtenbrot,\altaffilmark{2}
Scott~F.~Anderson,\altaffilmark{3}
W.~N.~Brandt,\altaffilmark{4, 5}
Aleksandar~M.~Diamond-Stanic,\altaffilmark{6}
Xiaohui~Fan,\altaffilmark{6}
Paulina~Lira,\altaffilmark{7}
Hagai~Netzer\,\altaffilmark{2}
Richard~M.~Plotkin,\altaffilmark{8}
Gordon~T.~Richards,\altaffilmark{9}
Donald~P.~Schneider,\altaffilmark{4}
and Michael~A.~Strauss\altaffilmark{10}
}

\altaffiltext{1}
               {Department of Physics, University of
               North Texas, Denton, TX 76203; ohad@unt.edu}
\altaffiltext{2}
                {School of Physics and Astronomy and the Wise Observatory, The Raymond
                and Beverly Sackler Faculty of Exact Sciences, Tel-Aviv University, Tel-Aviv
                69978, Israel}
\altaffiltext{3}
               {Department of Astronomy, University of Washington, Box 351580,
               Seattle, WA 98195}
\altaffiltext{4}
               {Department of Astronomy \& Astrophysics, The Pennsylvania State
               University, University Park, PA 16802}
\altaffiltext{5}
               {Institute for Gravitation and the Cosmos, The Pennsylvania State
               University, University Park, PA 16802}
\altaffiltext{6}
               {Steward Observatory, University of Arizona, 933 North Cherry
               Avenue, Tucson, AZ 85721}
\altaffiltext{7}
                 {Departamento de Astronom\'{i}a, Universidad de Chile, Camino del Observatorio
                 1515, Santiago, Chile}               
\altaffiltext{8}
                 {Astronomical Institute `Anton Pannekoek', University of Amsterdam,
                 Science Park 904, 1098 XH, Amsterdam, the Netherlands}
\altaffiltext{9}
               {Department of Physics, Drexel University, 3141 Chestnut Street,
               Philadelphia, PA 19104}
\altaffiltext{10}
               {Princeton University Observatory, Peyton Hall, Princeton,
               NJ 08544}

\begin{abstract}
 We present \geminiN\ $K$-band spectra of two representative members of the
 class of high-redshift quasars with exceptionally weak rest-frame ultraviolet emission lines (WLQs),
 \hbox{SDSS~J114153.34$+$021924.3} at \hbox{$z=3.55$} and \hbox{SDSS~J123743.08$+$630144.9} at \hbox{$z=3.49$}. In both sources  we detect an unusually weak broad \hb\ line and we place tight upper limits on the strengths of their [\ion{O}{3}] lines.
Virial, \hb-based black-hole mass determinations indicate normalized accretion rates of \hbox{\lledd$=0.4$} for these sources, which is well within the range observed for typical quasars with similar luminosities and redshifts.
We also present high-quality \xmm\ imaging spectroscopy of SDSS~J114153.34$+$021924.3 and find a hard-\xray\ photon index of
\hbox{$\Gamma=1.91^{+0.24}_{-0.22}$} which
supports the virial \lledd\ determination in this source. 
Our results suggest
that the weakness of the broad-emission lines in WLQs is not a consequence of an extreme continuum-emission source
but instead due to abnormal broad-emission line region\,properties.
 \end{abstract}


\section{The Nature of Lineless Quasars at High Redshift}
\label{introduction}

The Sloan Digital Sky Survey (SDSS; York \et 2000) has, so far, discovered $\sim80$
sources at \hbox{$2.2 \leq z \leq 5.9$} with \hbox{$\nu L_{\nu} (2500~\mbox{\AA}) \gtsim 10^{46}$\,ergs\,s$^{-1}$}
that have almost featureless rest-frame ultraviolet (UV) spectra (Fan \et 1999; Anderson \et 2001; Collinge \et 2005; Schneider \et 2005; Diamond-Stanic \et 2009, hereafter DS09; Plotkin \et 2010).
By virtue of their largely featureless spectra, the redshifts of these sources can be determined reliably only from the onset of the
Ly$\alpha$ forest or the presence of a Lyman limit system.
Subsequent  multiwavelength observations of several sources of this class have shown that they are unlikely to be high-redshift
galaxies with apparent  quasar-like luminosities due to gravitational lensing
amplification, dust-obscured quasars, or broad-absorption line quasars
(e.g., Shemmer \et 2006; DS09).
Furthermore, based on their relative radio and \xray\ weakness, Shemmer \et (2009, hereafter S09) have argued against the possibility that such sources may be the long-sought, high-redshift BL Lacertae objects (e.g., Stocke \& Perrenod 1981; see also DS09).
Instead, S09 suggested that these sources are unbeamed quasars with extreme properties.

High-redshift quasars with weak or undetectable UV emission lines (hereafter weak-emission line quasars, or WLQs) are
defined as quasars having rest-frame equivalent widths (EWs) of \hbox{$<15.4$~\AA} for the \hbox{Ly$\alpha +$\ion{N}{5}} emission-line complex (DS09). This EW threshold marks the $3\,\sigma$ limit at the low-EW end for the \hbox{Ly$\alpha +$\ion{N}{5}} EW distribution in a sample of $\sim3,000$ SDSS quasars at $z>3$ with no broad absorption lines.
Instead of the $\sim4$ quasars statistically expected to lie below this threshold (assuming the EW distribution is lognormal),
56 WLQs are included in that sample, and there is no corresponding excess of `strong-lined' quasars beyond the $3\,\sigma$ limit at the high-EW end (DS09). The reason for this excess of quasars with extremely weak UV emission lines is not yet understood.
Spectroscopic monitoring of four WLQs further suggests that it is unlikely that the weakness of their lines
can be explained by microlensing that temporarily and preferentially amplifies the continuum relative 
to the broad-emission lines (DS09). This result supports the idea that the UV emission lines in WLQs are intrinsically weak.

Intrinsically weak UV emission lines in quasars may be a consequence of a peculiar continuum-source spectral energy distribution (SED) that has a relative deficiency in high-energy photons. For example, a combination of large black-hole mass (\mbh) and extremely high accretion rate may result in a relatively narrow, UV-peaked SED (e.g., Leighly \et 2007a, b; Vasudevan \& Fabian 2007). In such a scenario, prominent high-ionization emission lines, such as \ion{C}{4}, are suppressed relative to low-ionization emission lines such as \hb, which are not affected to the same extent.
The peculiar spectral properties of the nearby \hbox{($z=0.19$)}, high-accretion rate quasar \hbox{PHL~1811}\ may be explained by this effect.
This quasar exhibits \hbox{EW(\ion{C}{4})$=6.6$\,\AA} (within the range observed for WLQs; DS09), while \hb,
for example, has a more typical, but still relatively weak, quasar value of \hbox{EW$=50$\,\AA} (Leighly \et 2007b).
At least one additional quasar at \hbox{$z<2.2$}, e.g., \hbox{PG~1407$+$265} (\hbox{$z=0.94$}; McDowell \et 1995),
may also be similar to \hbox{PHL~1811}.
Alternatively, WLQs may have abnormal broad-emission line region (BELR) properties such as a significant deficit of line-emitting gas in the BELR or a low BELR covering factor. In this scenario, their low-ionization emission lines are expected to be weak as well.
It has also been suggested that the BELRs in WLQs may be in an early stage of formation (Hryniewicz \et 2010).

\begin{deluxetable*}{lccccc}
\tablecolumns{6}
\tablewidth{0pc}
\tablecaption{\geminiN\ NIRI $K$-Band Observation Log}
\tablehead{
\colhead{} &
\colhead{} &
\colhead{} &
\colhead{$\log \nu L_{\nu}$} &
\colhead{} &
\colhead{Exp.~Time} \\
\colhead{Quasar} &
\colhead{$z$\tablenotemark{a}} &
\colhead{$z_{\rm sys}$\tablenotemark{b}} &
\colhead{$(2500$\,\AA)} &
\colhead{Observation Dates} & 
\colhead{(s)}
}
\startdata
\hbox{SDSS~J1141$+$0219} & 3.48 & 3.55 & 46.7 & 2009 Dec 12, 13, 23 &  6325 \\
\hbox{SDSS~J1237$+$6301} & 3.42 & 3.49 & 46.5 & 2009 Dec 23, 24, 25 &  10865
\enddata
\tablenotetext{a}{Redshift obtained from DS09, based on SDSS spectra.}
\tablenotetext{b}{Systemic redshift (see \S~\ref{NIRI_obs} for details).}
\label{log}
\end{deluxetable*}

\begin{deluxetable*}{lcccccccc}
\scriptsize
\tablecolumns{9}
\tablewidth{0pt}
\tablecaption{Derived Optical Properties}
\tablehead{ 
\colhead{} &
\colhead{} &
\multicolumn{3}{c}{Emission-Line Rest-Frame EW (\AA)} \\
\cline{3-5} &
\colhead{} &
\colhead{} &
\colhead{} &
\colhead{} \\
\colhead{Quasar} &
\colhead{FWHM(\hb)\tablenotemark{a}} &
\colhead{H$\beta$\,$\lambda4861$} &
\colhead{[\ion{O}{3}]\,$\lambda5007$} & 
\colhead{\ion{Fe}{2}\,$\lambda \lambda  4434-4684$\tablenotemark{b}} &
\colhead{$\alpha$\tablenotemark{c}} &
\colhead{$f_{\lambda}$\tablenotemark{d}} &
\colhead{$\log$ \mbh} &
\colhead{\lledd}
}
\startdata
\hbox{SDSS~J1141$+$0219} & $5.9^{+1.0}_{-1.1}$ & $20^{+4}_{-3}$ &{\phn} $\leq6$ & {\phn} $65^{+10}_{-9}$ & {\phn} $-2.15$ & {\phn} $1.07$ & 9.5 & 0.4 \\
\hbox{SDSS~J1237$+$6301} & $5.2^{+1.5}_{-1.0}$ & $35^{+8}_{-5}$ &{\phn} $\leq7$ & {\phn} $100^{+12}_{-13}$ & {\phn} $-0.98$ & {\phn} $0.63$ & 9.3 & 0.4
\enddata
\tablenotetext{a}{In units of 1000~\kms, corrected for instrumental resolution.}
\tablenotetext{b}{Broadened by a Gaussian profile having a FWHM of 2440~\kms\ and 1090~\kms, corrected for instrumental resolution, for \hbox{SDSS~J1141$+$0219} and \hbox{SDSS~J1237$+$6301}, respectively.}
\tablenotetext{c}{Optical continuum slope \hbox{($f_{\lambda} \propto \lambda^{\alpha}$)} determined in the 4700\,\AA$-$5100\,\AA\ rest-frame band.}
\tablenotetext{d}{Flux density at rest-frame 5100\,\AA\ in units of \hbox{$10^{-17}$\,ergs\,cm$^{-2}$\,s$^{-1}$\,\AA$^{-1}$}.}
\label{properties}
\end{deluxetable*}

In this work, we present the first near-infrared (NIR) spectroscopic observations of two representative WLQs,
\hbox{SDSS~J114153.34$+$021924.3} at \hbox{$z=3.55$} and \hbox{SDSS~J123743.08$+$630144.9} at \hbox{$z=3.49$}
(hereafter \hbox{SDSS~J1141$+$0219} and \hbox{SDSS~J1237$+$6301}, respectively).
Both sources were identified as candidate WLQs by Collinge \et (2005).
We also present a high-quality \xray\ spectrum of \hbox{SDSS~J1141$+$0219}, the first such of a WLQ.
These observations were designed as a pilot study to 1) determine the normalized accretion rates (in terms of the Eddington ratio, $L_{\rm bol}/L_{\rm Edd}$, hereafter \lledd) in WLQs and test the hypothesis that extremely high accretion rates are responsible for the weakness of the UV emission lines, and 2) measure the properties of low-ionization emission lines in WLQs.
We describe our observations and their results in \S\,\ref{observations}, discuss their implications for BELRs in active galactic nuclei (AGN) in \S\,\ref{discussion}, and provide a summary in \S\,\ref{summary}.
Throughout this {\em Letter}, luminosity distances are computed using the standard cosmological
model~(\hbox{$H_{0}=70$~\kms~Mpc$^{-1}$},~\hbox{$\Omega_{\Lambda}=0.7$},~\hbox{$\Omega_{M}=0.3$}).

\section{Observations, Data Reduction, and Results}
\label{observations}

\subsection{NIR Spectroscopy}
\label{NIRI_obs}

\begin{figure}
\plotone{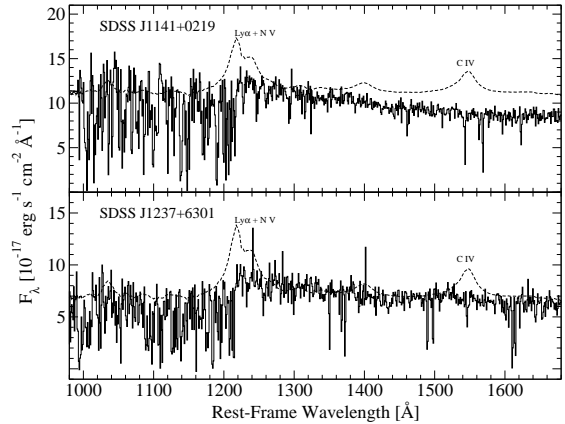}
 \caption{SDSS spectra of \hbox{SDSS~J1141$+$0219} ({\it top}) and \hbox{SDSS~J1237$+$6301} ({\it bottom}).
Both spectra were resampled in bins of 1\,\AA, for clarity. The SDSS Early Data Release quasar spectrum template (Vanden Berk \et 2001) is shown in each panel ({\it dashed lines}) for comparison; the SDSS quasar template was scaled arbitrarily in flux density, in each panel, and prominent emission lines are marked. Note the extreme weakness of the Ly$\alpha$ and \ion{C}{4} emission lines in each WLQ.}
\label{SDSS_spectra}
\end{figure}

\begin{figure}
\plotone{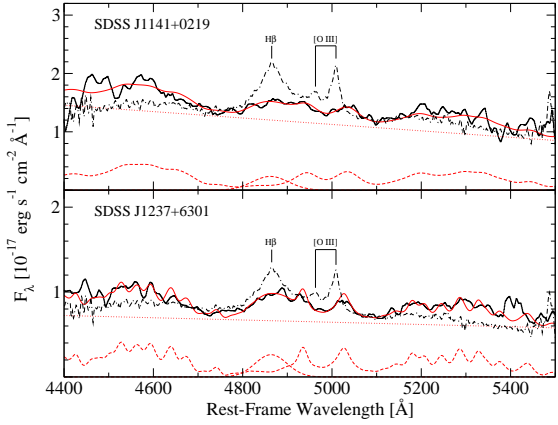}
 \caption{{\em Gemini-North} NIRI spectra of \hbox{SDSS~J1141$+$0219} ({\it top}) and \hbox{SDSS~J1237$+$6301}
 ({\it bottom}). Both spectra ({\it thick solid line}) were resampled in bins of 5\,\AA, for clarity. A mean spectrum of 33 quasars from the Shemmer \et (2004) and Netzer \et (2007) comparison sample is shown ({\it dot-dashed thick lines}) next to each WLQ spectrum for comparison; the mean spectrum of the comparison sample was scaled in flux to match the 5100\,\AA\ flux density in each WLQ spectrum, and prominent emission lines are marked. In each panel, the best-fit model ({\it red thin solid line}) is composed of
a continuum ({\it red dotted line}), and a broadened \ion{Fe}{2} emission complex and an \hb\ emission line ({\it red dashed lines}).
Note the extreme weakness of the \hb\ emission line in each WLQ.}
\label{NIRI_spectra}
\end{figure}

SDSS~J1141$+$0219 and \hbox{SDSS~J1237$+$6301}  were the brightest WLQs with suitable redshifts for observing the \hb\ spectral region
in the center of the $K$ band.\footnote{Prior to the Plotkin \et (2010) study, there were no suitable candidates for $H$-band spectroscopy, i.e., WLQs at \hbox{$2.2\ltsim z\ltsim 2.5$}.}
Their optical, \xray, and radio properties are typical of the WLQ population and, in particular, their \hbox{EW(Ly$\alpha$+\ion{N}{5})} are 3.3\,\AA\ and 4.6\,\AA, respectively (DS09; S09). The SDSS spectra of these WLQs are shown in Fig.~\ref{SDSS_spectra}.
Spectroscopic observations were carried out at the \geminiN\ Observatory using the Near InfraRed Imager and Spectrometer (NIRI; Hodapp \et 2003) as program \hbox{GN-2009A-Q-9}.
The observation log appears in Table~\ref{log}. For both targets, we used the $f/6$ camera with the 6-pixel centered slit (0.75$''$ wide), the $K$ grism, and the broad-band $K$ filter.
The resulting spectral range and spectral resolution were \hbox{$\Delta \lambda = 1.90-2.49 \, \mu {\rm m}$} and \hbox{$R=520$}, respectively.
Both targets were nodded along the slit to obtain optimal background subtraction. Exposure times for each subintegration were 275\,s and 265\,s and the total integration times were 6325\,s and 10865\,s for \hbox{SDSS~J1141$+$0219} and \hbox{SDSS~J1237$+$6301}, respectively.
We reduced the spectra using standard procedures of the {\sc IRAF}\footnote{IRAF (Image Reduction and Analysis Facility) is distributed by the National Optical Astronomy Observatory, which is operated by AURA, Inc.,
under cooperative agreement with the National Science Foundation.} {\sl Gemini} package (v\,1.8).
Exposures from different nodding positions were used to subtract the sky emission and were then co-added. 
Wavelength calibration employed Ar exposures. Relative flux calibration was obtained by taking spectra of telluric standard stars (A0 V type) immediately before or after the science exposures.
To compensate for possible slit losses in the flux calibration, we compared the total fluxes of our targets to those of nearby Two Micron All Sky Survey (2MASS; Skrutskie \et 2006) point sources, imaged in the NIRI \hbox{$2' \times2'$} field-of-view.
The uncertainty in the absolute flux calibration is estimated as $\sim10$\%, which is the quoted uncertainty on the 2MASS magnitudes.
The final NIR spectra appear in Fig.~\ref{NIRI_spectra}. The useful range of observed wavelengths is \hbox{$\sim 2.0-2.4~\mu$m}, corresponding to rest-frame \hbox{$\sim4400-5400$\,\AA}.

We modeled the spectra following the methods of Shemmer \et (2004). In short, our model consisted of
a linear continuum, fitted between two narrow ($\pm20$\,\AA) rest-frame bands centered on 4700\,\AA\ and 5100\,\AA,
a broadened \ion{Fe}{2} emission template (Boroson \& Green 1992), a broad \hbox{(1,200\,\kms\,$\leq$\,FWHM\,$\leq$\,15,000\,\kms)} Gaussian profile, representing \hb, and three narrow \hbox{(300\,\kms\,$\leq$\,FWHM\,$\leq$\,1,200\,\kms)} Gaussian profiles: one for the narrow \hb\ component and two for \hbox{[\ion{O}{3}]\,$\lambda\lambda$\,4959,\,5007}. The widths of the narrow Gaussians were tied to the same value and the [\ion{O}{3}] lines were constrained to have the theoretical ratio \hbox{$I($[\ion{O}{3}]\,$\lambda5007)/I($[\ion{O}{3}]\,$\lambda4959)=2.95$}.
The \ion{Fe}{2} emission template was broadened by convolving it with a Gaussian profile whose width was free to vary within the velocity range used for the broad \hb.
Our fitting results are summarized in Table~\ref{properties}, where error bars are given at a 90\% confidence level, and the best-fit models appear in Fig.\,\ref{NIRI_spectra}.
Upper limits on \hbox{EW([\ion{O}{3}]\,$\lambda5007$)} were determined by assuming a Gaussian profile with \hbox{FWHM$=1000$~\kms} embedded in random noise at the observed signal-to-noise level ($\sim10$) and looking for the weakest such feature that would have been detected in our spectra (see, e.g., Shemmer \et 2004).

In both sources we detect unusually weak broad \hb\ lines as well as prominent \ion{Fe}{2} emission (cf. Netzer \& Trakhtenbrot 2007). 
Since [\ion{O}{3}] lines are undetected, we consider the centroids of the best-fit \hb\ lines for determining the systemic redshifts of the sources (Table~\ref{log}).
These redshifts are larger by \hbox{$\Delta z=0.07$} than those determined from the SDSS spectra (DS09). This indicates that the SDSS redshift determination may be biased towards somewhat lower values, since the onset of the Ly$\alpha$ forest is not easily detectable at \hbox{$z\sim3.5$}. The rest-frame spectra in Fig.~\ref{SDSS_spectra} were corrected by $z_{\rm sys}$.
We note that the broadening of the \ion{Fe}{2} templates required Gaussian profiles that are significantly narrower
than those of their corresponding \hb\ lines (Table~\ref{properties}), which is not
uncommon if most of the \ion{Fe}{2} emission is considered to originate in an `intermediate-line region'
(e.g., Hu \et 2008). 
Fig.~\ref{NIRI_spectra} also shows the average \hb\ spectral region of 33 radio-quiet quasars (RQQs) at \hbox{$2.0 \ltsim z \ltsim 3.5$} without broad absorption lines
(hereafter, the comparison sample) from Shemmer \et (2004) and Netzer \et (2007)
whose luminosities
are comparable to those of our WLQs.
The sources from the comparison sample all have \hbox{EW(\hb)$>$46\,\AA}; the maximum is \hbox{EW(\hb)$=256$\,\AA} and the mean is \hbox{EW(\hb)$=107$\,\AA}. Thus, our two WLQs lie at the tail of an EW(\hb) distribution for high-redshift quasars,
although this result is not as significant as the deviation of WLQs from the lognormal distribution of \hbox{EW(Ly$\alpha$+\ion{N}{5})} in high-redshift quasars (\S\,\ref{introduction}).

\subsection{The X-ray Spectrum of \hbox{SDSS~J1141$+$0219}}
\label{xmm_obs}

We obtained \xray\ imaging spectroscopy of \hbox{SDSS~J1141$+$0219} with
\xmm\ (Jansen \et 2001) on 2008 June 27 (dataset \hbox{ID~0551750301}).
The data were processed using standard \xmm\ Science Analysis
System\footnote{http://xmm.esac.esa.int/sas} v8.0.0 tasks.
The event files were filtered in time to remove periods of
flaring activity in which the count rates of each MOS (pn) detector exceeded
0.35 (1.0) \hbox{counts~s$^{-1}$} for events having \hbox{$E>10$\,keV}; the net
exposure times were 19.7~ks and 16.7~ks for the MOS1/MOS2 and pn
detectors, respectively.
For each detector, source counts were extracted using a circular aperture with
\hbox{$r=30''$} centered on the source, and background counts were
extracted from a nearby source-free region that was at least as large as
the corresponding source region.
The spectrum from each detector was grouped with a minimum of 20 counts per bin.
The net source counts in the \hbox{0.2--10.0\,keV} band were 81, 128, and 333 for the MOS1, MOS2, and pn
detectors, respectively.

\begin{figure}
\epsscale{1.1}
\plotone{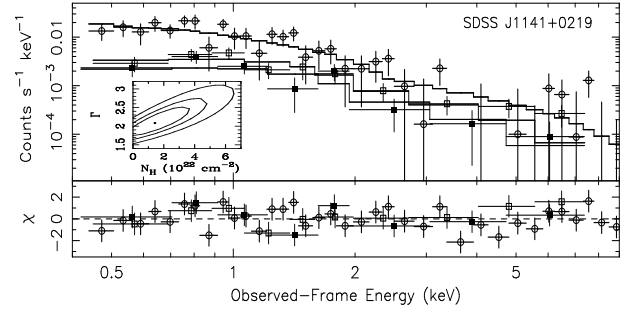}
 \caption{The \xmm\ spectrum of \hbox{SDSS~J1141$+$0219}.
 Open circles, filled squares, and open squares represent the EPIC pn, MOS1, and MOS2 detector data, respectively. Solid lines
 represent the best-fit model for each spectrum, and the thick lines mark the best-fit model for the pn data. The data were fitted with a
 Galactic-absorbed power-law model at observed-frame energies \hbox{$>$0.44\,keV}. The $\chi$ residuals are in units of $\sigma$ with error bars of size~1. The inset shows 68\%, 90\%, and 99\% confidence contours for the photon index ($\Gamma$) and an intrinsic, neutral-absorption column density (\nh), when the latter component is added to the Galactic-absorbed power-law model.}
\label{XMM_spectrum}
\end{figure}

Joint spectral fitting of the data from all three detectors was performed with \hbox{XSPEC v11.3.2} (Arnaud 1996),
considering only observed-frame energies above 0.44\,keV. This energy range corresponds
to  rest-frame energies above 2\,keV, where the underlying power-law hard \xray\ spectrum is less prone to
contamination due to any potential absorption or soft excess emission.
We fitted the spectrum with a power-law model and a Galactic-absorption component, which was kept fixed during the
fit with a column density of \hbox{\nh$=2.30 \times 10^{20}$\,cm$^{-2}$} obtained from Dickey \& Lockman (1990).
The power-law model consisted of a single photon index for
all three detectors, while the power-law normalizations were free to vary.
The best-fit photon index is \hbox{$\Gamma=1.91^{+0.24}_{-0.22}$} (\hbox{$\chi^2=45.2$} for 48 degrees of freedom; at 90\% confidence).
The \xray\ spectrum and its best-fit model and residuals appear in Figure~\ref{XMM_spectrum}.
Based on an \hbox{$F$-test}, we find that adding an intrinsic, neutral-absorption component to the above model 
does not improve the fit significantly and hence is not warranted by our data.
The upper limit on the column density of such intrinsic neutral-absorption component is \hbox{\nh $\leq3.59 \times10^{22}$\,cm$^{-2}$}
(at 90\% confidence).
Finally, we do not detect \xray\ variability between the \xmm\ and shallow \chandra\ observations of this source reported in S09.

\section{Extremely High Accretion Rates or Anemic Broad-Line Regions?}
\label{discussion}

We find that at least two WLQs exhibit extremely weak UV as well as \hb\ emission lines that place them at the
tail of the corresponding EW distributions in type~1 AGN.
On the other hand, a wealth of multiwavelength observations of a sub-set of WLQs have
shown that, unlike their optical-UV emission-line properties, their broad-band continua are consistent
with those observed in the general quasar population at the corresponding redshifts and
luminosities (e.g., DS09, S09).
Such observations already question the leading hypothesis that WLQs are quasars with extremely high accretion rates (i.e., \hbox{\lledd$\gtsim1$}), which has been motivated mainly by the peculiar properties of \hbox{PHL~1811}\ (Leighly \et 2007a,b; \S\,\ref{introduction}).
This quasar shows unusually weak UV emission lines and it has \hbox{\lledd\,$\sim 1.3$}, derived from an \hb-based, virial determination of \mbh.
The properties of \hbox{PHL~1811}\ can also be viewed as an extreme case of the anticorrelation between UV-line (e.g., \ion{C}{4}) strength and \lledd\ (Baskin \& Laor 2004).
Based on the very small \ion{C}{4} EWs (or EW upper limits) of WLQs (DS09), and assuming this anticorrelation extends to their relatively high luminosities, the implication is \hbox{\lledd$\gtsim1$} for all WLQs.
Such a range of \lledd\ constitutes the high-end tail of the \lledd\ distribution in
$\sim60,000$ SDSS quasars (Shen \et 2008), and this may be analogous to the excess of WLQs
with respect to the lognormal distribution of UV-line EWs in high-redshift quasars (\S~\ref{introduction}).

We determine virial \mbh\ and corresponding \lledd\ values for our WLQs from
their derived optical properties (Tables~\ref{log} and \ref{properties}),
utilizing the empirical BELR size-luminosity relation of Kaspi \et (2005) modified by Bentz \et (2009).
This results in the following expressions for \mbh\ and \lledd:

\begin{equation}
\label{eq:mbh}
\frac{M_{\rm BH}}{10^{6}M_{\sun}}=5.05\,\left[\frac{\nu
    L_{\nu}(5100\,\mbox{\AA})}{10^{44}\,{\rm
      ergs\,s^{-1}}}\right]^{0.5}\left[\frac{{\rm FWHM}({\rm
      H}\beta)}{10^3\,{\rm km\,s^{-1}}}\right]^2
\end{equation}

\begin{equation}
\label{eq:lledd}
L/L_{\rm Edd}=0.13f(L)\left[\frac{\nu L_{\nu}(5100\,\mbox{\AA})}{10^{44}\,{\rm ergs\,s^{-1}}}\right]^{0.5}\left[\frac{{\rm FWHM}({\rm
      H}\beta)}{10^3\,{\rm km\,s^{-1}}}\right]^{-2},
\end{equation}
where we have employed Equation~21 of Marconi \et (2004) to obtain
$f(L)$, the luminosity-dependent bolometric correction to
\hbox{$\nu L_{\nu}(5100\,\mbox{\AA})$}, which is 5.7 for both sources.
The resulting \mbh\ and \lledd\ values (Table~\ref{properties}) are well within the ranges observed in recent quasar surveys (e.g., Shen \et 2008), and there are no indications for extremely high accretion rates in our WLQs.
In other words, although the \mbh\ uncertainty associated with a virial-\mbh\ determination is typically on the order of a factor of $\sim2$, given the luminosities of our WLQs, \hbox{\lledd$>1$} would imply \hbox{FWHM(\hb)$<3500$\,km\,s$^{-1}$} (according to
Eq.~\ref{eq:lledd}), which is significantly lower than observed.
The weakness of the \hb\ lines in our WLQs, resulting in unusually high \hbox{$I($\ion{Fe}{2}$)/I($\hb$)$} ratios in these sources (thus implying extremely high \lledd\ values; Table~\ref{properties}), may appear inconsistent with the rather `normal' \lledd\ values we derive (e.g., Netzer \et 2004). However, we note that several sources, mainly at \hbox{$z\ltsim0.75$}, are known to have high
\hbox{$I($\ion{Fe}{2}$)/I($\hb$)$} ratios and yet have \hbox{\lledd\,$\sim0.1$} (e.g., Netzer \& Trakhtenbrot 2007). It is therefore possible that our WLQs constitute high-redshift outliers from the empirical \hbox{EW(\hb)-\lledd} anticorrelation of Netzer \et (2004).
We note that the absence of [\ion{O}{3}] emission lines in our WLQs is consistent with their high \hbox{$I($\ion{Fe}{2}$)/I($\hb$)$} ratios;
furthermore, $\sim20$\% of high-redshift quasars have [\ion{O}{3}] lines that are as weak as those we measure\,(Netzer \et 2004).

Another accretion rate indicator in luminous quasars is the slope of the hard-\xray\ spectrum (e.g., Shemmer \et 2008; Risaliti \et 2009). For \hbox{SDSS~J1141$+$0219}, we measure a hard-\xray\ photon index of \hbox{$\Gamma=1.91^{+0.24}_{-0.22}$}, which is typical of luminous, high-redshift RQQs. This measurement corresponds to \hbox{\lledd\,$\sim0.6$}, based on the empirical \hbox{$\Gamma$-\lledd} correlation (Shemmer \et 2008), corrected for the BELR size-luminosity relation used here. \hbox{PHL~1811}, for comparison, has \hbox{$\Gamma=2.28^{+0.12}_{-0.11}$} (Leighly \et 2007a), consistent within the errors, with the prediction from this correlation given its \hbox{\lledd\,$\sim1.3$}.
However, a more pronounced difference between \hbox{PHL~1811}\ and \hbox{SDSS~J1141$+$0219} lies in their \hbox{optical--\xray} SEDs, since the former has been found to be \hbox{$\approx30-100$} times \xray\ weaker than expected from its optical luminosity (corresponding to \hbox{$\approx 3-5~\sigma$} deviation; Steffen \et 2006; Gibson \et 2008) while the latter is \xray\ {\em brighter} than expected by a factor of
$\sim3$ (corresponding to \hbox{$\sim1.3~\sigma$} deviation; see S09). The extreme \xray\ weakness of \hbox{PHL~1811}\ has been attributed to its extremely high \lledd\ (Leighly \et 2007a).
We find no \xray\ indication for unusually high \lledd\ in \hbox{SDSS~J1141$+$0219}, consistent with the virial-\mbh\ determination.
This result is consistent with the S09 findings in which there is no evidence of an exceptionally high photon index in a joint \xray\ spectral fit of~11~WLQs.

Based on the \hbox{optical--UV} spectra of two WLQs, we find that the weakness of their emission lines
is not a consequence of unusual continuum source properties.
Alternatively, our results suggest that the weakness of both low- and high-ionization potential BELR lines in our two WLQs may be a consequence of unusual BELR physical properties.
For example, weak (or `anemic') BELR lines across the spectrum may be due to gas deficit in the BELR manifested by a low BELR covering factor.
We note that `anemic' emission-line regions in AGNs have also been suggested to be a result of exceptionally {\em low} \lledd\  (e.g., Nicastro 2000), but this scenario does not apply to WLQs that are significantly more luminous than the so called `optically-dull AGN' found mainly in the local Universe (e.g., Trump \et 2009).
The key to understanding the BELR in WLQs is to investigate the relationship between their high- and low-ionization BELR lines (e.g., \ion{C}{4} and \hb) by additional NIR spectroscopy (to measure low-ionization lines) coupled with detailed photoionization modeling.

\section{Summary}
\label{summary}

We present new $K$-band spectra of two WLQs at \hbox{$z\sim3.5$} and an \xmm\ spectrum of one of them. The $K$-band spectra allowed measurement of the spectral properties of the \hb\ lines and the \ion{Fe}{2} emission complex, and the placement of tight upper limits on the strengths of the [\ion{O}{3}] lines. The broad \hb\ lines in both sources are significantly weaker than those observed in typical quasars with similar luminosities and redshifts, but they still enable reliable determinations of \mbh\ and \lledd. The results of our \hbox{$K$-band} and \xray\ spectra do not support the idea that the weakness of the high- as well as low-ionization BELR lines in our two WLQs can be accounted for by extremely high accretion rates. Instead, our results are suggestive of `anemic' BELRs in WLQs.
Clearly, high-quality NIR and \xray\ spectroscopy of additional WLQs are required for robust \lledd\ determinations in order to test this possibility further in a much larger sample.

\acknowledgments

We~gratefully~acknowledge the financial support of~NASA grants \hbox{NNX09AF04G} (O.\,S) and \hbox{NNX10AC99G} (W.\,N.\,B), and NSF grant \hbox{AST-0707266} (M.\,A.\,S). We thank D.~Chelouche, M.~Gaskell, and B.~Wills for fruitful~discussions.

This work is based on observations obtained at the Gemini Observatory, which is operated by the Association of Universities for Research in Astronomy, Inc., under a cooperative agreement with the NSF on behalf of the Gemini partnership: the National Science Foundation (United States), the Science and Technology Facilities Council (United Kingdom), the National Research Council (Canada), CONICYT (Chile), the Australian Research Council (Australia), Minist\'{e}rio da Ci\^{e}ncia e Tecnologia (Brazil) and Ministerio de Ciencia, Tecnolog\'{i}a e Innovaci\'{o}n Productiva (Argentina). This work is also based on observations obtained with \xmm, an ESA science mission with instruments and contributions directly funded by ESA Member States and NASA.

Funding for the SDSS and SDSS-II has been provided by the Alfred P. Sloan Foundation, the Participating Institutions, the National Science Foundation, the U.S. Department of Energy, the National Aeronautics and Space Administration, the Japanese Monbukagakusho, the Max Planck Society, and the Higher Education Funding Council for England. The SDSS Web Site is
http://www.sdss.org/.

\end{document}